\documentclass[12pt]{article}
\usepackage{amsfonts,amssymb}
\usepackage{enumerate}
\usepackage{verbatim}
\usepackage{graphicx}
\usepackage{epsf}
\usepackage{subfig}
\usepackage{cleveref}
\usepackage{color}
\usepackage{transparent}

\usepackage{fullpage}

\begin{document}

\title{On Bogomolny-Schmit conjecture }

\author{D Beliaev and Z Kereta}

\maketitle

\begin{abstract}
%In \cite{BSch02} 
Bogomolny and Schmit proposed that the critical edge percolation on the square lattice is a good model for the nodal domains of a random plane wave. Based on this they made a conjecture about the number of nodal domains. Recent computer experiments showed that the mean number of clusters per vertex and the mean number of nodal domains per unit area are very close but different. Since the original argument was mostly supported by numerics, it was believed that the percolation model is wrong. In this paper we give some numerical evidence in favour of the percolation model.
\end{abstract}

\section{Introduction}
We start by fixing the definitions. The random plane wave  could be defined as
$$
\phi(z)=\sum_{n=-\infty}^\infty C_n J_{|n|}(kr)e^{i n\theta}
$$
where $C_n$ are i.i.d. Gaussians subject to the condition $C_{-n}=\overline{C_n}$ (this guaranties that the function is real). One can think that this function is a Gaussian vector in the $L^2$ space of real solutions of Helmholtz equation $(\Delta+k^2)f=0$. %Similarly one can consider a random Gaussian vector in the space of spherical harmonics of degree $n$. 
This random function is of special interest since it is conjectured to be a good universal description of high-energy Laplace eigenfunctions in domains with chaotic dynamics \cite{Berry77}.
We are interested in the behaviour of the nodal lines and nodal sets of function $\phi$.

Many local questions such as what is the average length of the nodal lines per unit area can be answered using Kac-Rice techniques \cite{Berry02,Wigman09}. The global questions are much more involved. One of the quantities of interest is the mean number of connected components (per unit area) \cite{BSch02,BGS}.  

Behaviour of the random plane wave is very closely related to that of a random spherical harmonic. Random spherical harmonic of degree $n$ is a Gaussian vector in the space of all spherical harmonics of degree $n$. It is known that after proper rescaling random spherical harmonics converge to random plane waves. It is known \cite{NaSo09} that  for  a random spherical harmonic of degree $n$ the number of the nodal domains normalized by $n$ is exponentially concentrated around some strictly positive constant. The proof of this result is local, hence the same is true if we consider nodal domains restricted to a fixed open sub-domain of the sphere.  Moreover, the same argument should imply that the result is true for the random plane waves and the number of nodal components per unit area is the same.

Bogomolny and Schmit argued that the nodal lines can in the {\em mean} be considered as forming a rectangular lattice with $\mathrm{Area}(\Omega) k^2/(2\pi^2)$ sites. This normalization constant can be obtained from the mean spacing between nodal lines or de Broglie wave length $\lambda=2\pi/k$. 
Thus one can think that there are two square lattices with $\mathrm{Area}(\Omega) /\lambda^2$ sites that are dual to each other and nodal lattice is the medial lattice between them.  Based on this analogy and explicit computations for the percolation on the square lattice they conjectured that the mean number of nodal domains scales as
\begin{equation}
\label{eq:BS}
 \frac{\mathrm{Area}(\Omega)k^2}{4\pi}\, \frac{3\sqrt{3}-5}{\pi}\approx\frac{\mathrm{Area}(\Omega)k^2}{4\pi}\, 0.0624373.
\end{equation}
This claim was supported by numerical simulations on a relatively small scale. 

For historical reasons we will use this normalization and will study the number of nodal domains divided by $A k^2/4\pi$ (by Weyl's law this is the number of eigenvalues of Laplacian in $\Omega$ that are below $k^2$). Alternatively, (\ref{eq:BS}) can be rewritten as
$$
2\, \frac{\mathrm{Area}(\Omega)k^2}{4\pi^2}\, \frac{3\sqrt{3}-5}{2}
$$ 
were the last factor is the mean number of bond percolation clusters per site (for the square lattice). In other words, this is the mean number of primal and dual bond clusters for a square lattice containing $\mathrm{Area}(\Omega)/\lambda^2$ sites \cite{ZFA}. 

\section{Results}

\subsection*{Number of nodal domains}
Since the work of Bogomolny and Schmit there were several computer experiments that computed this density on a larger scale. The first result is by M.~Nastasescu \cite{Nastasescu11} who computed the mean number of nodal domains for random spherical harmonics (this is a spherical analogue of the random plane wave; the densities are the same for both models). She found that the density is $0.0598\pm 0.0003$ which is $5\sigma$ away from the Bogomolny-Schmit conjecture. Originally, this discrepancy was attributed to finite size and curvature effects.   
In \cite{Konrad12} it was shown that the density for plane wave  is $0.0589\pm 0.000142$.  It is believed that the normalized number of nodal domains behaves like $a+b/k$. In \cite{Konrad12} the number of nodal domains was computed for $100$ samples for $k=100,\dots,1100$. The best fit to the data was $0.0589+4.6209/k$.

We performed computer simulations  with different values of $k$. For sampling of the random plane wave we used a code developed by A.~Barnett. Given a sample we compute the number of connected component using standard clustering algorithms. The main difficulty in this computation is resolving the so-called ``nearly-avoided crossings''. It is easy to show that the nodal lines can not intersect. Indeed this would imply that at some point simultaneously $\phi$, $\phi_x$ and $\phi_y$ are equal to zero. These three functions are independent Gaussians, hence the probability of this event is zero. On the other hand, two different nodal lines can be arbitrarily close to each other and on a discrete level this creates an ambiguity when we have to determine the connectivity. We resolve these ambiguities using local interpolation.  Further technical details can be found in  \cite{Kereta12}.

\begin{figure}[h]
        \centering
        \includegraphics[width=0.9\textwidth]{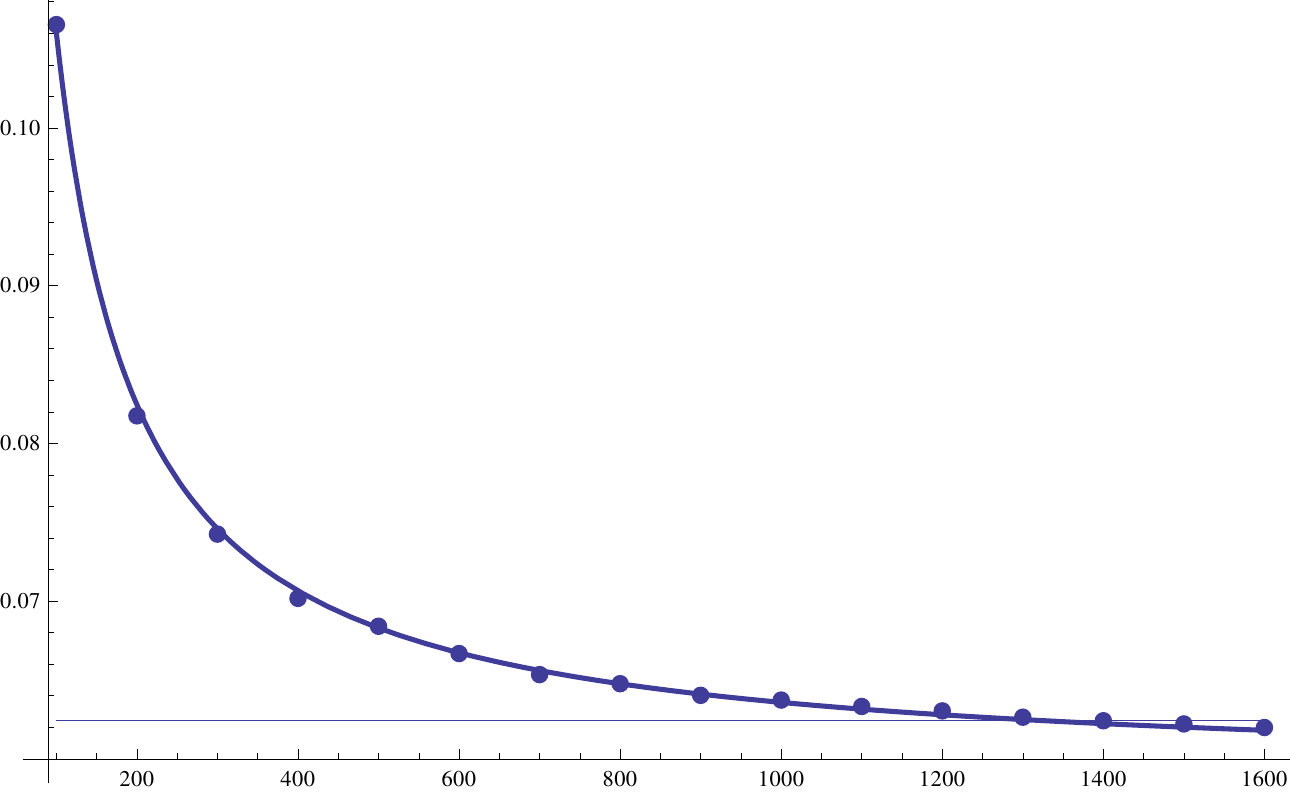}
        \caption{Normalized mean number of nodal domains as a function of $k$. The horizontal lines gives the Bogomolny-Schmit prediction.}
	\label{fig:density}
\end{figure}

The results of the simulations are given by Figure  \ref{fig:density}. We computed the normalized number of nodal domains for $k=100,\dots,1600$ and $100$ samples for each $k$. The best fit to the data is given by $a+b/k$ with $a=0.0589$ and $b=4.717$. The first constant is in good agreement with results of \cite{Konrad12}. The second constant is two orders of magnitude larger than $a$, this explains why one has to go to very high energies to recover the correct value of $a$.

The results of Nastasescu, Konrad, and our simulations give strong numerical evidence that the Bogomolny-Schmit prediction is slightly off. 

{\em Remark.} Bogomolny and Schmit used the percolation model where the number of the lattice points was chosen according to the wavelength. Namely they used the square lattice with edge-length equal to the wavelength. Another possible normalization is to use the square lattice such that the number of vertices is the same as the number of local maxima of the random plane wave. The average density of the  critical points can be computed using Gaussian integrals and is equal to $k^2 0.0919$ and one quarter of them are maxima. Using this normalization, the percolation model gives us the following estimate for the number of nodal domains
$$
\frac{\mathrm{Area}(\Omega)k^2}{4\pi}\,0.0566.
$$
This is a bit closer to the numerical results, but still too far to be  a good model.

\subsection*{Crossing probability}

We want to make an important remark:  it is believed that the density of nodal domains is a universal quantity in the following sense. Let $M$ be a compact manifold and let $\psi_n$ be the eigenfunctions of Laplace operator. We define $f_n=\sum_{n}^{n+C\sqrt{n}} c_k \psi_k$ where $c_k$ are i.i.d normal variables and $C$ is a large constant depending on the metric. Then the properly rescaled number of nodal domains of $f_n$ has asymptotically the same density as the random plane wave. On the other hand, the number of clusters per vertex is a non-universal quantity in percolation theory, it strongly depends on the lattice structure. 

Another important remark is that essentially all microscopic observables for percolation are lattice dependent, but the macroscopic picture is conjectured to have conformally invariant scaling limit which is universal and described by Schramm-Loewner evolution with parameter $\kappa=6$ (this proved only for site percolation on triangular lattice \cite{Smirnov01}).   This suggests to study an observable which is macroscopic and hence universal from percolation theory point of view. The simplest observable to consider is the crossing probability. In this paper we consider two crossing events. Let $\Omega$ be a rectangle with side lengths $\lambda$ and $1$ then we define $P_h^k(\lambda)$ to be the probability that there is a nodal line of a random plane wave with energy $k^2$ which connects left and right sides of $\Omega$ (inside $\Omega$). The second observable is $P_{hv}^k(\lambda)$, the probability that there is a nodal line connecting all four sides of $\Omega$. 

The percolation counterparts are probabilities that there is an open cluster connecting sides of $\lambda N \times N$ rectangle. It is known \cite{Cardy92,Watts96} that as $N\to\infty$ these probabilities converge to $P_{h}$ and $P_{hv}$ that are given by
\begin{eqnarray}
P_{h}(\lambda)&=&\frac{\Gamma(2/3)}{\Gamma(1/3)\Gamma(4/3)}(1-m)^{1/3}{}_2F_1\left(\frac{1}{3},\frac{2}{3},\frac{4}{3},1-m\right),\\
P_{hv}(\lambda)&=&P_{h}(\lambda)-\frac{(1-m)}{\Gamma(1/3)\Gamma(2/3)}{}_3F_2\left(1,1,\frac{4}{3},2,\frac{5}{3},1-m\right),
\end{eqnarray}
where ${}_2F_1$ and ${}_3F_2$ are hypergeometric functions and
$$
m=\frac{\theta_4^4(0,e^{-\pi/\lambda})}{\theta_3^4(0,e^{-\pi/\lambda})}
$$
where $\theta_3$ and $\theta_4$ are Jacobi theta functions.

Using the same numerical techniques as for the number of nodal domains we computed the probabilities of horizontal and simultaneous horizontal and vertical crossings. We computed these probabilities for $\lambda$ from $1$ to $3$ with step $0.2$. In all cases we used $k=1000$ which corresponds to the rectangle whose longest side is equal to $159$ wave-lengths.

\begin{figure}[h] 
\begin{tabular}{p{7.2cm}p{7.2cm}}
\includegraphics[width=7.2cm]{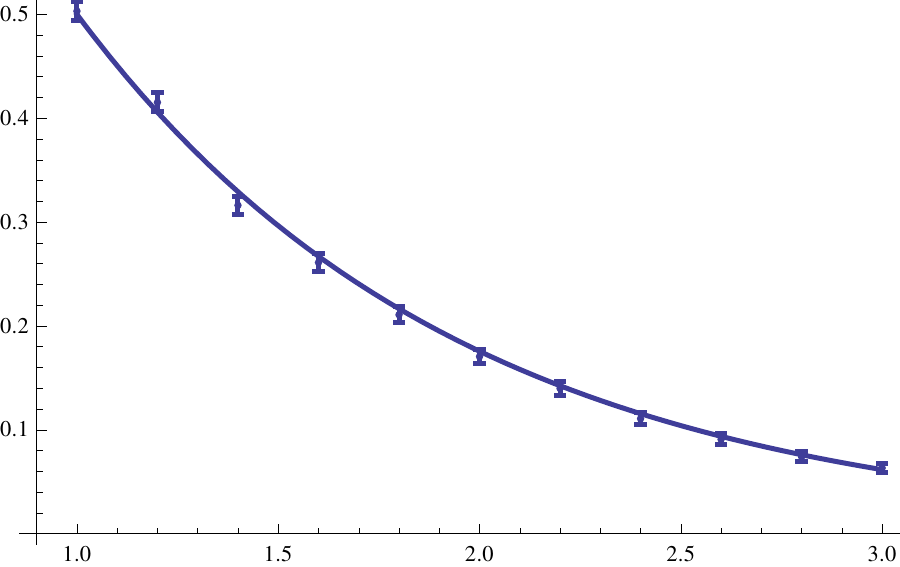}&
%\hspace{-1cm}
\includegraphics[width=7.2cm]{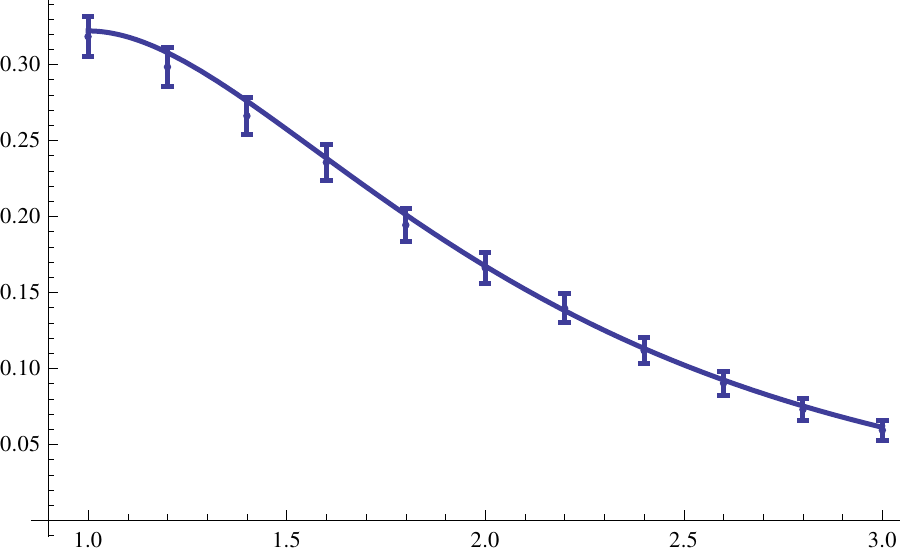}
\end{tabular}
\caption{\label{fig:cardy}\label{fig:watts}
Probability of having a left-right crossing (left) and of having both left-right and top-bottom crossings (right) as functions of $\lambda$. Solid lines are Cardy's and Watts' formulas. In both cases the longest side of the rectangle is $1$ and $k=1000$. We used $10000$ samples for each value of $\lambda$ for the horizontal  crossing and $5000$ samples for both crossings.
}
\end{figure}

Figures \ref{fig:cardy}  show the data and $95\%$ confidence intervals for both types of crossing. This gives a very strong indication that the crossing probabilities for random plane wave converge to Cardy's and Watts' formulas.  Important observation is that the random plane wave is rotationally invariant, hence this result is a bit stronger than the similar result for percolation. Its percolation analog would be crossing probability for a rectangle arbitrarily rotated with respect to the lattice. It is also known \cite{Smirnov01} that if crossing probability for arbitrary conformal rectangle (i.e. simply-connected domain with two marked disjoint boundary arcs) with conformal modulus $\lambda$ converges as $k\to\infty$ to $P_h(\lambda)$, then we have that macroscopic nodal lines have conformally invariant scaling limit which is given by SLE($6$).

{\em Remark.} There are other approaches to conformal invariance of the scaling limit: Harris criterion \cite{BSch07} or direct comparison with SLE curves \cite{BDSch07}.

\section*{Conclusion}
 Numerical results give very strong numerical evidence that conjecture (\ref{eq:BS}) is wrong, the correct density is $0.0589$ instead of $0.0624$, and the critical bond percolation on the square lattice is not a good model for local observables. On the other hand macroscopic observables match very well which suggests that the nodal lines have conformally invariant scaling limit which is described by conformal field theory with $c=0$ or Schramm-Loewner Evolution with $\kappa=6$.

\section*{Acknowledgement} We would like to thank A.~Barnett for allowing us to use his code for sampling the random plane waves. We also would like to thank the referees for their comments and for bringing some of the related results to our attention.

\bibliographystyle{iopart-num}
\bibliography{nodal} 
\end{document}